%% file: main.tex
\documentclass[aps,prl,reprint]{revtex4-2} 

\usepackage[english]{babel}
\usepackage[utf8]{inputenc}
\usepackage{amsmath, amssymb, amsthm}
\usepackage[pdftex, pdftitle={Article}, pdfauthor={Author}]{hyperref} 
\usepackage{xcolor}
\usepackage{xspace}
\input{MB_commands.tex}

\newcommand{\field}[0]{\phi}
\newcommand{\rfield}[0]{\widetilde{\phi}}
\newcommand{\vbar}[0]{\bar{v}}
\newcommand{\her}[2]{\text{H}_{#1}\left( #2 \right)}

\newcommand{\elabel}[1]{\label{eqn:#1}}
\newcommand{\action}{\mathcal{A}}
\newcommand{\actionPert}{\action_{\text{pert}}}
\newcommand{\eref}[1]{(\ref{eqn:#1})}
\newcommand{\Eref}[1]{Eq.~\eref{#1}}
\newcommand{\Erefs}[1]{Eqs.~\eref{#1}}
\newcommand{\imag}{\mathring{\imath}}

\newcommand{\latin}[1]{{\it #1}}
\newcommand{\ie}{\latin{i.e.}\@\xspace}

\usepackage{amsmath}
\usepackage{accents}
\usepackage{mathtools}

\DeclareMathSymbol{\widetildesym}{\mathord}{largesymbols}{"65}
    \newcommand\lowerwidetildesym{%
      \text{\smash{\raisebox{-1.3ex}{%
        $\widetildesym$}}}}

\newcommand\parwidetilde[1]{%
   \mathchoice
      {\accentset{\displaystyle\scalebox{.3}{(}\lowerwidetildesym\scalebox{.3}{)}}{#1}}
         {\accentset{\textstyle\scalebox{.3}{(}\lowerwidetildesym\scalebox{.3}{)}}{#1}}
         {\accentset{\scriptstyle\scalebox{.3}{(}\lowerwidetildesym\scalebox{.3}{)}}{#1}}
         {\accentset{\scriptscriptstyle\scalebox{.3}{(}\lowerwidetildesym\scalebox{.3}{)}}{#1}}
   }

\newcommand{\fieldpar}{\parwidetilde{\field}}

\begin{document}
\newcommand{\titleText}{Field Theory of free Active Ornstein-Uhlenbeck Particles}
\title{\titleText}

\author{Marius Bothe}
    \email[Correspondence email address: ]{mpb19@ic.ac.uk}
    \affiliation{Imperial College London}
\author{Gunnar Pruessner}
    \email[Correspondence email address: ]{g.pruessner@imperial.ac.uk}
    \affiliation{Imperial College London}

\date{\today} 

\begin{abstract}
We derive a Doi-Peliti field theory for free active Ornstein-Uhlenbeck particles, or, equivalently, free inertial Brownian particles, and present a way to diagonalise the Gaussian part of the action and calculate the propagator. Unlike previous coarse-grained approaches this formulation correctly tracks particle identity and yet can easily be expanded to include potentials and arbitrary reactions.
\end{abstract}


\maketitle

\section{Introduction}
The field of active matter \cite{marchetti_hydrodynamics_2013} is concerned with systems whose microscopic constituents draw energy from internal processes, which allows them to act independently from external forces, such as self-propulsion in a medium.

This is realized in many biological systems, where organisms metabolize chemical fuel to enable their locomotion, with examples as diverse as moving bacteria
\cite{berg_e_nodate},
desert locusts 
\cite{buhl_disorder_2006},
schools of fish
\cite{hubbard_model_2004}
and cell migration
\cite{trepat_physical_2009}.
Beyond examples from biology there are several artificial realisations of active matter systems
\cite{paxton_catalytic_2004,volpe_microswimmers_2011,deseigne_collective_2010,van_der_linden_interrupted_2019},
with potential applications in nano- and micromachines.

Theoretically there have been several mesoscopic approaches to modeling the phenomenon of active matter, notably active Brownian particles and run-and-tumble particles, for example \cite{fily_athermal_2012} and \cite{schnitzer_theory_1993} respectively. However the model of interest for the present work is the active Ornstein-Uhlenbeck (AOU) process: Here we consider an overdamped particle moved by its own self-propulsion velocity, $v$, which is driven by a Gaussian noise and restrained by a friction force. As reviewed in \cite{martin_statistical_2020} this process is defined by two coupled differential equations,
\begin{subequations}
\label{eqn:AOUPmodel}
\begin{align}
\label{eqn:AOUPmodelPosition}
\der{}{t}\ve{r}(t) &= - \mu \nabla V(\ve{r}) + \ve{v}(t) + \sqrt{2D_x}\ve{\eta}_x(t)\,, \\
\tau \der{}{t} \ve{v}(t) &= - \ve{v}(t) + \sqrt{2D} \ve{\eta}_v(t) \,,
\elabel{AOUPmodelVelocity}
\end{align}
\end{subequations}
where $V(\ve{r})$ is the potential, $v(t)$ is the self-propulsion velocity of the particle at time $t$, $\mu$ is a coupling regulating the strength of the potential, and $1/\tau$ is the friction coefficient, an inverse time. The noises $\eta_x(t)$ and $\eta_v(t)$ are Gaussian, with correlation function $\langle \eta_i(t) \eta_j(t')\rangle =\delta(t-t') \delta_{i,j}$. While in the literature it is common to allow for noise only in velocity $\ve{v}$, we consider the more general case here, of having noise and thus diffusion also in the position $\ve{r}$. The expressions derived are continuous in the spatial diffusion constant $D_X$, and the case of no noise in the position is indeed recovered by setting $D_x=0$. This model has received a lot of attention\cite{martin_statistical_2020}, and was recently shown, in a course grained description, to display a motility-induced phase separation \cite{fodor_how_2016, maggi_universality_2020} with a repulsive potential, a feature of active matter systems. However there is still no systematic field-theoretic description of active Ornstein-Uhlenbeck particles so far, despite considerable interest \cite{nardini_entropy_2017,stenhammar_continuum_2013}. The present work fills this gap.

The description of an AOU particle with $D_x=0$ is closely related to the equations of motion for an inertial Brownian particle:  
\begin{subequations}
\label{eqn:inertialParticle}
\begin{align}
\der{}{t}\ve{r} &= \ve{v} \,, \\
\tau \der{}{t} \ve{v} &= - \ve{v} - \mu' \nabla V(\ve{r}) + (2D)^{1/2} \ve{\eta}_v. \label{eqn:inertialParticleVelocity}
\end{align}
\end{subequations}
The difference between the systems (\ref{eqn:AOUPmodel}) and (\ref{eqn:inertialParticle}) lies entirely in the fact that in the inertial Brownian case (\ref{eqn:inertialParticle}) the force $\nabla V$ acts on the velocity (\ref{eqn:inertialParticleVelocity}), rather than on the position $\ve{r}$, (\ref{eqn:AOUPmodelPosition}), as in the AOU process. This is a consequence of the different interpretations of the velocity variable in both systems: For the AOU particles $\ve{v}$ is not just the derivative of the position, but rather an active self-propulsion velocity, which undergoes an Ornstein-Uhlenbeck process completely independently from the potential $V$. AOU particles and inertial Brownian particles are indistinguishable in the absence of a potential, and they are not active \cite{fodor_how_2016}. The framework derived here describes both free AOU particles as well as inertial Brownian particles, and to the authors knowledge constitutes their first field-theoretic description that takes into account particle identity. This is a crucial step for a systematic expansion around non-linearities that may constitute activity. Furthermore the Doi-Peliti formalism used here can easily be adapted to include arbitrary reactions \cite{pausch_topics_2019}, for example a birth-death process, making this approach very versatile.

%


\section{Deriving the Field Theory}
In the following we will restrict the discussion of the field theory to the one-dimensional case for easier readability. A generalization to higher dimensions is straightforward as the required expressions are simply sums or products of those in one dimension. 

The quickest way to arrive at a field-theoretical description of the AOU process that respects the particle nature of the constituents is to move from the Langevin \Eref{inertialParticle} to a Fokker-Planck equation, following the standard procedure \cite{risken_fokker_1996}:
\begin{widetext}
\begin{align}
\elabel{FPE}
\pder{}{t} p(\ve{r},\ve{v},t) &= \left[ D_x \nabla^2_r + \frac{D}{\tau^2} \nabla^2_v - \ve{v} \nabla_r  + \mu \nabla_r V'(\ve{r}) + \frac{1}{\tau} \nabla_v \cdot \vec{v} \right] p(\ve{r},\vec{v},t) 
\qquad\text{with}\qquad V'(x)=\nabla_r V(\ve{r}) \,.
\end{align}
Following the approach in \cite{Garcia-MillanPruessner:ToBePublished}, this is directly turned into a Doi-Peliti action\cite{doi_second_1976,peliti_path_1985} of the form $\action=\action_0+\actionPert$, with
\begin{subequations}
\begin{align}
\mathcal{A}_0 &= \int \diff x \diff v \diff t 
\ \rfield(x,v,t) \left[ \pder{}{t} - D_x \npder{}{x}{2} - \frac{D}{\tau^2} \npder{}{v}{2} + v \pder{}{x} - \frac{1}{\tau} \pder{}{v} v  \ \right] \field(x,v,t) \,, \label{eqn:action}
\\ 
\actionPert &= - \int \diff x \diff v \diff t
\ \rfield(x,v,t) \mu \pder{}{x} \left(
V'(x) \field(x,v,t) \right) 
\qquad\text{with}\qquad V'(x)=\frac{\mathrm{d}V(x)}{\mathrm{d}x} \,,
\end{align}
\end{subequations}
\end{widetext}
in one dimension. All operators in the square brackets above act on the entire product to their right, including $p$ and $\field$ respectively.
The external potential $V(x)$ in \Eref{AOUPmodelPosition} has been moved to the perturbative part of the action, as it is usually not easily integrated, except for special cases, such as a harmonic one, similar to the effective friction potential of the velocity here \cite{WalterPruessnerSalbreux:2020,Garcia-MillanPruessner:2020}. 

More traditionally, one can recast the process \eref{AOUPmodel} in a master equation on a lattice \cite{TaeuberHowardVollmayr-Lee:2005,pausch_topics_2019} and obtain \Eref{action} after taking the continuum limit on the resulting action. This is done in detail in the supplementary material, see \Eref{hamiltonian}. Expectations of the fields are then calculated as
\begin{equation}
    \mean{\bullet} = 
    \int \Diff{\field} \Diff{\rfield} 
    e^{-\action_0-\actionPert}
    = \mean{\bullet\,e^{-\actionPert}}_0 \,,
\end{equation}
so that
\begin{equation}
    \mean{\bullet}_0 = 
    \int \Diff{\field} \Diff{\rfield} 
    e^{-\action_0} \ .
\end{equation}
As with any perturbative field theory, the key-ingredient is the propagator
$\mean{\field(x_1,v_1,t_1)\rfield(x_0,v_0,t_0)}_0$
of the bilinear action $\action_0$, which is the focus of the remainder of this work.

The alternative path towards a field-theory of 
using the much more popular Martin-Siggia-Rose "trick" \cite{martin_statistical_1973,janssen_lagrangean_1976,de_dominicis_field-theory_1978} requires subsequent mending of the particle nature by additional terms \cite{Dean1996} that can be difficult to handle.

\section{Diagonalizing the Action}
Once the action is diagonalised, it is easy to read off the propagators using 
\begin{equation}
    \int \diff z^*\!\!\wedge\!\diff z\ \exp{-z^*z a} z^{*n} z^m =2\pi\imag a^{-(n+m+1)} n! \delta_{n,m} \ ,
\end{equation}
for any $a$ with positive real part.
In the following we diagonlise the action $\action_0$ \Eref{action} in three steps. Firstly, we  Fourier transform in both $t$ and $x$: 
\begin{subequations}
\elabel{def_Fourier}
\begin{align}
\fieldpar(x,v,t) &= 
\int \dbar{\omega} 
\dbar{k} \exp{\imag \vec{k} \vec{x}-\imag \omega t} \fieldpar(k,v,\omega)\,,\\
\fieldpar(k,v,\omega) &= 
\int \diff t \diff x \exp{-\imag \vec{k} \vec{x}+\imag \omega t} \fieldpar(x,v,t) \,,
\end{align}
\end{subequations}
where any powers of $2\pi$ have been absorbed into the measure $\dbar k = \diff k /(2\pi)$ and similar for $\dbar \omega$. We will use a similar convention for the $\delta$-function, $\deltabar = 2\pi \delta$. 


Using \Eref{def_Fourier} for both $\field$ and $\rfield$ we obtain for the Gaussian part of the action \eref{action}:
\begin{widetext}
\begin{equation}\elabel{action_after_Fourier}
\mathcal{A}_0 = \int \dbar k \dbar \omega \diff v \rfield(-k,v,-\omega) \left[ -\imag \omega +D_x k^2- \frac{D}{\tau^2} \npder{}{v}{2} + \imag vk - \frac{1}{\tau} \pder{}{v} v \right] \field(k,v,\omega) \,.
\end{equation}
Because of the harmonic potential  term 
$\tau^{-1} \partial_v (v \field)$ and the mixed term $\imag v k$, we cannot diagonalise the action by simply Fourier transforming in $v$ as well. Instead we follow similar considerations for the harmonic potential acting on a particle's position using Hermite polynomials \cite{WalterPruessnerSalbreux:2020,Garcia-MillanPruessner:2020} and the eigenvalue problem of the Kramer's equation \cite{risken_fokker_1996}. 

As a second step we therefore rescale the fields by an exponential: 
\begin{subequations}
\elabel{field_rescaling}
\begin{align}
\field(k,v,\omega) = \exp{-\frac{\tau}{4D} v^2} \field^{\prime}(k,v,\omega) , \\
\rfield(k,v,\omega) = \exp{\frac{\tau}{4D} v^2} \rfield^{\prime}(k,v,\omega),
\end{align}
\end{subequations}
resulting in the transformed action
\begin{align}
\mathcal{A}_0 = \int \dbar k \dbar \omega \diff v \rfield^{\prime}(-k,v,-\omega) \left[ -\imag \omega +D_x k^2  
- \frac{D}{\tau^2} \npder{}{v}{2} 
+ \imag vk 
+ \frac{v^2}{4D} -\frac{1}{2\tau}\right]  \field^{\prime}(k,v,\omega) \label{eqn:action2}\,.
\end{align} 
In this expression the term $\imag v k$ couples $v$ and $k$, which left untreated would require a perturbative expansion. However,
one can complete the square in $v$ by introducing the shifted variable $\vbar(k,v)=v+2\imag Dk$, so that 
$\partial_{\vbar}=\partial_v$ and 
$\imag k v + v^2/(4D)= D k^2 + \vbar^2/(4D)$, thereby decoupling $v$ and $k$.
In the following we will sometimes omit the explicit dependence on $k$ by simply writing $\vbar$, however it is important to keep it in mind when considering multiple fields $\field^{\prime}(\omega,k,v)$, whose individual $\vbar(k,v)$ depend on different $k$ and $v$. 
The shift yields
\begin{align}
\mathcal{A}_0 = \int \dbar k \dbar \omega \diff v \rfield^{\prime}(-k,v,-\omega) \left[ -\imag \omega + (D_x+D) k^2 - \frac{D}{\tau^2} \npder{}{\vbar}{2} + \frac{\vbar^2}{4D} -\frac{1}{2 \tau} \right] \field^{\prime}(k,v,\omega)  \; . 
\label{eqn:action1}
\end{align}
Having removed the coupling between $v$ and $k$, the resulting operator in the action has the form of a quantum harmonic oscillator in $\vbar$ \cite{nolting2017theoretical}.
The eigenfunctions of this operator can be written elegantly in terms of Hermite polynomials  $\her{n}{x}$. In particular:
\begin{align}
\begin{split}
&\left[ -i \omega + (D_x+D)k^2 - \frac{D}{\tau^2} \npder{}{\vbar}{2}  + \frac{1}{4D} \vbar^2 -\frac{1}{2 \tau} \right] 
\exp{-\frac{\tau}{4D}\vbar^2}  \her{n}{\vbar\sqrt{\frac{\tau}{2D}}}\\
& \quad =\exp{-\frac{\tau}{4D}\vbar^2}  \left[ -i \omega + (D_x+D) k^2 - \frac{D}{\tau^2} \npder{}{\vbar}{2} +\frac{1}{\tau} \vbar \pder{}{\vbar} \right]
\her{n}{\vbar\sqrt{\frac{\tau}{2D}}}  \,,
\end{split} \\
& \quad = \exp{-\frac{\tau}{4D}\vbar^2}  \left[ -i \omega + (D_x+D) k^2 +\frac{n}{\tau}  \right]  \her{n}{\vbar\sqrt{\frac{\tau}{2D}}} \; ,
\label{eqn:eigenvalueHermite}
\end{align}
where in the last line we used the eigenvalue equation of the Hermite polynomials \cite{courant_methods_1975},
$(\partial_z^2 - 2 z \partial_z ) \her{n}{z} = -2n \her{n}{z}$,
which obey the orthogonality and the completeness relations
\begin{equation}
\int_{-\infty}^{\infty} \diff \vbar 
\frac{\exp{-\vbar^2/V^2}}{2^n n! \sqrt{\pi}}
\her{n}{\frac{\vbar}{V}} \her{m}{\frac{\vbar}{V}}  = V \delta_{n,m}
\qquad
\text{ and }
\qquad
    \frac{1}{V} \sum_{n=0}^\infty
    \frac{\exp{-\vbar^2/V^2}}{2^n n! \sqrt{\pi}}
    \her{n}{\frac{\vbar}{V}} \her{n}{\frac{\vbar'}{V}}
    = \delta(\vbar-\vbar') \ .
\end{equation}
for arbitrary positive $V$, in particular in the present context $V=\sqrt{2D/\tau}$.
To make use of this simplification we introduce one more transformation, namely
\begin{subequations}
\elabel{Hermite_transform}
\begin{align}
    \field'(k,v,\omega) & = 
    \exp{-\frac{\tau}{4D} \vbar^2(k,v)}
    \sqrt{\frac{\tau}{2D}} \sum_{n=0}^\infty 
    \her{n}{\vbar(k,v) \sqrt{\frac{\tau}{2D}} } \field_n(k,\omega)\\
    \rfield'(k,v,\omega) & = 
    \exp{-\frac{\tau}{4D} \vbar^2(-k,v)}
    \sqrt{\frac{\tau}{2D}} \sum_{n=0}^\infty 
    \frac{1}{2^n n! \sqrt{\pi}} 
    \her{n}{\vbar(-k,v)  \sqrt{\frac{\tau}{2D}} } \rfield_n(k,\omega)
\end{align}
\end{subequations}
Combining the transformations \Erefs{Hermite_transform}, \eref{field_rescaling} and \eref{def_Fourier} amount to
\begin{subequations}
\label{eqn:transform}
\begin{align}
\field(x,v,t) &= \int \dbar \omega \dbar k \exp{\imag kx -\imag \omega t} \exp{-\frac{\tau}{4D} v^2} \exp{-\frac{\tau}{4D} \vbar^2(k,v)} \sqrt{\frac{\tau}{2D}} \sum_{n=0}^{\infty} \her{n}{\vbar(k,v)\sqrt{\frac{\tau}{2D}}} \field_n(k,w) \; , \\
\rfield(x,v,t) &= \int \dbar \omega \dbar k \exp{\imag kx -\imag \omega t} \exp{\frac{\tau}{4D} v^2} \exp{-\frac{\tau}{4D} \vbar^2(-k,v)} \sqrt{\frac{\tau}{2D}} \sum_{n=0}^{\infty} \frac{1}{2^n n!\sqrt{\pi}} \her{n}{\vbar(-k,v)\sqrt{\frac{\tau}{2D}}} \rfield_n(k,w) \; ,
\end{align}
\end{subequations}
where position $x$ and time $t$ are Fourier transformed into the continuous variables $\omega$ and $k$ and the velocity $v$ is effectively replaced by the discrete Hermite index $n$. Any sum over this discrete index is preceded by a factor of $\sqrt{\tau/(2D)}$ to simplify dimensional consistency. To maintain orthogonality in Hermite indices, the Hermite transform of the annihilator field $\field(\omega,k,v)$ at $k$ is based on $\vbar(k,v)$, while the Hermite transform of the creator field $\rfield(-\omega,-k,v)$ multiplying it, \Eref{action_after_Fourier}, is based on $\vbar(-k,v)$. 
\Erefs{transform} relate the transformed fields $\fieldpar_n(k,w)$ directly to the original fields $\fieldpar(x,v,t)$ of the original action \Eref{action}. 
The corresponding inverse transforms read 
\begin{subequations}
\label{eqn:itransforms}
\begin{align}
\field_n(k,\omega) = \int \diff x \diff t \diff v \exp{-\imag kx+\imag \omega t} \exp{\frac{\tau}{4D}v^2} \exp{-\frac{\tau}{4D}\vbar^2(k,v)} \frac{1}{2^n n!\sqrt{\pi}} \her{n}{\vbar(k,v)\sqrt{\frac{\tau}{2D}}} \field(x,v,t) \; , \\
\rfield_n(k,\omega) = \int \diff x \diff t \diff v \exp{-\imag kx+\imag \omega t} \exp{-\frac{\tau}{4D}v^2} \exp{-\frac{\tau}{4D}\vbar^2(-k,v)} \her{n}{\vbar(-k,v)\sqrt{\frac{\tau}{2D}}} \rfield(x,v,t) \; .
\end{align}
\end{subequations}
Using the unitary transform \Eref{Hermite_transform} in the action \Eref{action1} gives
\begin{align}
\mathcal{A}_0 = \int \dbar k \dbar \omega 
\sqrt{\frac{\tau}{2D}}
\sum_{n=0}^{\infty} \rfield_n(-k,-\omega) \left[ -i \omega  + (D_x + D) k^2 +\frac{n}{\tau} \right] \field_n(k,\omega) \; ,
\end{align}
\end{widetext}
so that the bare propagator of this theory can be read off from the action to be
\begin{equation}\elabel{propagator}
    \mean{\field_n(k,\omega) \rfield_m(k',\omega')} =
    \frac{\deltabar(\omega+\omega') \deltabar(k+k')\sqrt{\frac{2D}{\tau}} \delta_{n,m}}{-\imag\omega + (D_x+D) k^2 + n/\tau} \,.
\end{equation}
The $\delta$-functions on the right indicate translational invariance in time and space, as well as an equivalence of the Hermite operator acting on the final velocity or its adjoint on the initial. In other words, if \Eref{propagator} obeys the Fokker-Planck equation on the final $v$, it equally obeys the backward Kolmogorov equation on the initial $v'$.
The term $(D_x+D) k^2$ indicates a mean squared displacement of $2(D_x+D)t$ asymptotically in large $t$, \ie as far as mean squared displacement is concerned, $D_x$ and $D$ in \Eref{AOUPmodel} are additive. However, $D$ enters independently in the transform \Eref{transform}, where it is not shifted by $D_x$. The index of the Hermite polynomials now plays the role of the mass through the term $\frac{n}{\tau}$ in denominator.

None of the above relies on $D_x \neq 0$, and so we conclude that the propagator \eref{propagator} is well-defined, non-singular and continuous in $D \leq 0$ provided $D_x > 0$.


\section{The Propagator}
In the following we express the propagator $\langle \field(x_1,v_1,t_1) \rfield(x_0,v_0,t_0) \rangle$, which is the probability density of finding a particle at position $x_1$ with velocity $v_1$ at time $t_1$, having placed it initially at position $x_0$ with velocity $v_0$ at time $t_0$. It is the inverse transform \eref{Hermite_transform} of the propagator \eref{propagator}. The Doi-Peliti formalism and its commutator guarantee that all correlation functions respect the particle nature of the degrees of freedom. We find
\begin{widetext}
\begin{align}
&\mean{\field(x_1,v_1,t_1)\rfield(x_0,v_0,t_0)}_0  
= \int \dbar{\omega_1} \dbar{\omega_0} \dbar{k_1} \dbar{k_0} 
\exp{\imag k_1 x_1 + \imag k_0 x_0} \exp{-\imag  \omega_1 t_1 -\imag  \omega_0 t_0} \exp{-\frac{\tau}{4D}(v_1^2-v_0^2+\vbar^2(k_1,v_1) + \vbar^2(-k_0,v_0))} \\
\nonumber &\qquad \times \frac{\tau}{2D} \sum_{n_1,n_0=0}^{\infty} 
\frac{1}{2^{n_0}n_0!\sqrt{\pi}} 
\her{n_0}{\vbar(-k_0,v_0) \sqrt{\frac{\tau}{2D}}} 
\her{n_1}{\vbar(k_1,v_1)\sqrt{\frac{\tau}{2D}}} 
\frac{\deltabar(\omega_1+\omega_0) \deltabar(k_1+k_0)\sqrt{\frac{2D}{\tau}} \delta_{n_1,n_0}}{-\imag\omega_1 + (D_x+D) k_1^2 + n_1/\tau}\\
\nonumber =&   \int \dbar{k_1} \exp{\imag k_1 (x_1-x_0)}  
\exp{-\frac{\tau}{4D}(v_1^2-v_0^2+\vbar^2(k_1,v_1) + \vbar^2(k_1,v_0))} 
\sqrt{\frac{\tau}{2D}}
\sum_{n_1=0}^{\infty} 
\frac{1}{2^{n_1}n_1!\sqrt{\pi}}
\her{n_1}{\vbar(k_1,v_0) \sqrt{\frac{\tau}{2D}}} 
\her{n_1}{\vbar(k_1,v_1) \sqrt{\frac{\tau}{2D}}} \\
\nonumber&\qquad \times \int \dbar{\omega_1} \frac{\exp{-\imag \omega_1(t_1 - t_0)}}{-\imag  \omega_1 + (D_x+D)k_1^2 +n_1/\tau}\\
\nonumber =& \int \dbar{k_1} \exp{\imag k_1 (x_1-x_0)}  
\exp{-\frac{\tau}{4D}(v_1^2-v_0^2+\vbar^2(k_1,v_1) + \vbar^2(k_1,v_0))} 
\exp{-(D_x+D)k_1^2(t_1-t_0)}\\
\nonumber & \qquad \times
\sqrt{\frac{\tau}{2D}}
\sum_{n_1=0}^{\infty} 
\frac{1}{2^{n_1}n_1!\sqrt{\pi}}
\her{n_1}{\vbar(k_1,v_0) \sqrt{\frac{\tau}{2D}}} 
\her{n_1}{\vbar(k_1,v_1) \sqrt{\frac{\tau}{2D}}} 
\left(\exp{-(t_1-t_0)/\tau}\right)^{n_1}
\end{align}
To simplify the expression we introduce $t=t_1-t_0$ and Mehler's formula \cite{MagnusOberhettingerSoni:1966}:
\begin{align}
\sum_{n=0}^{\infty} \frac{1}{2^n n!} H_n(x) H_n(y) z^n = \frac{1}{\sqrt{1-z^2}} \exp{\frac{-x^2 z^2 + 2 xyz - y^2 z^2}{1-z^2}} \; .
\end{align}
Using this we find for the propagator
\begin{align}
& \mean{\field(x_1,v_1,t_1)\rfield(x_0,v_0,t_0)}_0 
= 
\int \dbar{k_1} \exp{\imag k_1 (x_1-x_0)}  
\exp{-\frac{\tau}{4D}(v_1^2-v_0^2+\vbar^2(k_1,v_1) + \vbar^2(k_1,v_0))} 
\exp{-(D_x+D)k_1^2(t_1-t_0)}\\
\nonumber & \qquad \times
\sqrt{\frac{\tau}{2\pi D}}
\frac{1}{\sqrt{1-\exp{-2t/\tau}}} 
  \EXP{ - \frac{\tau}{2D} 
\frac{ \big( \vbar^2(k_1,v_0) + \vbar^2(k_1,v_1) \big) \exp{-2t/\tau} 
- 2 \vbar(k_1,v_0) \vbar(k_1,v_1) \exp{-t/\tau}}{1- \exp{-2t/\tau}} }
\end{align}
By substituting $\vbar(k_1,v_{0,1})=v_{0,1} +2 \imag D k_1 $ and performing the Gaussian integral in $k_1$ we arrive at the final result:
\begin{align}
& \mean{\field(x_1,v_1,t_1)\rfield(x_0,v_0,t_0)}_0 
= \sqrt{\frac{\tau}{2 \pi D (1-\exp{-2t/\tau})}} \EXP{- \frac{\tau}{2D} \frac{(v_1-v_0 \exp{-t/\tau})^2}{1- \exp{-2t/\tau}}} \\
& \qquad \times  \frac{1}{\sqrt{4 \pi D \left( t-2\tau \left( \frac{1-\exp{-t/\tau}}{1+\exp{-t/\tau} }\right)\right)+4 \pi D_x t}} \EXP{ - \frac{\left( x_1-x_0 -\tau (v_1+v_0)  \frac{1-\exp{-t/\tau}}{1+\exp{-t/\tau} } \right)^2 }{4D\left( t - 2 \tau \frac{1-\exp{-t/\tau}}{1+\exp{-t/\tau} }\right) + 4 D_x t} } \nonumber \; .
\end{align}
\end{widetext}
From the first exponential on the right hand side, we read off that the velocity distribution is initially centered at $v_0$, but this memory of the initial condition decays exponentially  with characteristic time $\tau$, the inverse friction. The distribution also widens as time passes, and approaches its maximum variance, $D/\tau$, asymptotically. Despite worrying competition between $t$ and $\tau$, the variance of the velocity is always strictly non-negative.

For the position variable, we see in the seond exponential that the center of the $x_1$ distribution is shifted by $x_0$ as well as a velocity dependent term. This term term scales linearly with time for small $t$, 
\begin{align}
\tau (v_1 + v_0) \frac{1-\exp{-t/\tau}}{1+ \exp{-t/\tau}} = \frac{v_1+v_0}{2} t + O(t^2) \,,
\end{align}
capturing the motion of the particle with both initial velocity $v_0$ and final velocity $v_1$, and approaching a constant offset for large $t$. Unlike the velocity, the position never fully loses its dependence on the initial $x_0$ and $v_0$, but its variance keeps growing linearly for large $t$. We further see that the diffusion in $x$ contributes additively to a widening of the Gaussian for the position, and has no impact on the velocity. The diffusion parameterised by $D_x$ may be thought of as being superimposed on the diffusion by random acceleration. In the limit $D_x \to 0$ the result agrees exactly with the propagator of the Kramer's equation \cite{risken_fokker_1996}. 

While this result was derived in one dimension, higher dimensions amount to adding analogous terms to the action. The propagator in real space and direct time will therefore factorise into a product of exponential functions, so that the d-dimensional density is just the product of one-dimensional densities.

\section{Conclusion}
The transformations found in this paper diagonalise the Gaussian part of the AOU action, resulting in a compact propagator in Fourier-Hermite space. This is a fundamental step, which opens the door for a field theoretical treatment of interacting AOU particles and at the same time gives a particle-identity respecting field theory of inertial Brownian particle.

Using this framework we found an expression for the propagator function that confirms our physical intuitions about the process, which is indistinguishable from a non-overdamped description of a free diffusive particle.

Future work building on this could be an investigation of motility induced phase separation \cite{cates_motility-induced_2015} in a field-theoretic framework, for example via volume exclusion or by introducing a repulsive pair potential. The fact that AOU particles can be cast in a Doi-Peliti field theory means it can easily be extended to include arbitrary reactions, such as birth-death processes, which could be the topic of further study.

\section*{Acknowledgements}

The authors would like to thank Mike Cates, Rosalba Garcia-Millan, Irene Li and Benjamin Walter for the interesting discussions.

\bibliographystyle{apsrev4-2}
\bibliography{bibliography}

\widetext
\vspace*{1cm}
\begin{center}
\textbf{\large Supplemental Materials: \titleText}
\end{center}
\setcounter{equation}{0}
\setcounter{figure}{0}
\setcounter{table}{0}
\setcounter{page}{1}
\renewcommand{\thepage}{S-\arabic{page}}
\makeatletter
\renewcommand{\theequation}{S\arabic{equation}}
\renewcommand{\thefigure}{S\arabic{figure}}
\renewcommand{\thetable}{S\arabic{table}}
\setcounter{section}{0}
\renewcommand{\thesection}{S-\Roman{section}}
\renewcommand{\thesubsection}{\arabic{subsection}}
\counterwithin*{equation}{section}
\counterwithin*{figure}{section}

\section{\label{sec:DP}Doi-Peliti Derivation}
In the following we derive a Doi-Peliti field theory via a master equation, which in turn is derived from the process in the Langevin \Erefs{AOUPmodel}. To this end, we will discretise the regular space of positions, as well as the space of velocities.

The master equation is written for a probability $p_t(\{n\})$ of finding the system at time $t$ in a certain occupation number state $\{n\}$, which is the vector of all occupation numbers $n_{x,v}$ of all phase space sites $x,v$. To simplify the continuum limit to be taken later, we discretise positions by a spacing $\Delta x$ and velocities by spacings of $\Delta v$.

A master equation has the general form
\begin{equation}
\elabel{master}
    \der{}{t} p_t(\{n\}) = \sum_{\{n'\}\ne \{n\}} 
    W(\{n'\} \to \{n\}) p_t(\{n'\})
    -
    \sum_{\{n'\}\ne \{n\}} 
    W(\{n\} \to \{n'\}) p_t(\{n\}) \,,
\end{equation}
where $W(\{n\} \to \{n'\})$ is the transition rate from occupation number state $\{n\}$ to $\{n'\}$. The transitions being concurrent Poisson processes that change the occupation numbers, the rates may be non-zero only where $\{n'\}$ differs from $\{n\}$ in exactly two occupation numbers $n_{x,v}$ at nearest neighbouring sites.

\Erefs{AOUPmodel} describes two concurrent processes, namely a drift-diffusion in $x$, \Eref{AOUPmodelPosition}, and an Ornstein-Uhlenbeck process in $v$, \Eref{AOUPmodelVelocity}. Those processes may be superimposed in a master equation, writing $W(\{n\} \to \{n'\})=W_x(\{n\} \to \{n'\})+W_v(\{n\} \to \{n'\})$, where $W_x$ refers to the spatial process, and $W_v$ to the change in the velocity occupation.

To ease notation, we may specify $\{n'\}$ only as far as those occupation numbers are concerned that are different to $\{n\}$. The spatial hopping in $x$ is then characterised by the two hopping rates
\begin{subequations}
\begin{align}
W_x(\{n\} \to \{n_{x,v}-1, n_{x+\Delta x,v}+1\}) &= r_x(v) n_{x,v}\,,\\
W_x(\{n\} \to \{n_{x-\Delta x,v}+1, n_{x,v}-1\}) &= \ell_x(v) n_{x,v}\,,
\end{align}
\end{subequations}
to the right and to the left respectively.
Here $r_x(v)$ and $\ell_x(v)$ refer to the rates with which particles hop to the right and the left respectively given a certain velocity $v$. 
Normally, the continuum limit corresponds to taking $\Delta x\to 0$ with the constraints $\Delta x^2 (r_x + \ell_x)=2D_x$ and $\Delta x (r_x-\ell_x)=v$ constant determining the hopping rates $r_x$ and $\ell_x$ \cite{CocconiETAL:2020}.
Writing, correspondingly $r_x(v)=D_x/\Delta x^2 + v/(2\Delta x)$ may then result in negative hopping rates as the velocity $v$ is not constant, in principle not bounded and subject to evolution itself. To avoid negative rates we thus write
\begin{subequations}
\elabel{rates_hopping}
\begin{align}
r_x(v)    & = \frac{D_x}{\Delta x^2} + \frac{v \theta(v)}{\Delta x} \,,\\
\ell_x(v) & = \frac{D_x}{\Delta x^2} + \frac{- v \theta(-v)}{\Delta x} \,,
\end{align}
\end{subequations}
recovering $r_x-\ell_x = v (\theta(v) + \theta(-v))$.

The hopping in velocity space is complicated by the friction term that features like a harmonic potential \cite{Garcia-MillanPruessner:2020}, which provides the "spring-constant" of the potential and rescales the diffusive term, so that
\begin{subequations}
\begin{align}
W_v(\{n\} \to \{n_{x,v}-1, n_{x,v+\Delta v}+1 \}) &= r_v(v) n_{x,v}\,,\\
W_v(\{n\} \to \{n_{x,v-\Delta v}+1, n_{x,v}-1\}) &= \ell_v(v) n_{x,v}\,,
\end{align}
\end{subequations}
with
\begin{subequations}
\elabel{rates_v}
\begin{align}
r_v(v)    & = \frac{D}{\tau^2 \Delta v^2} - \frac{v}{2 \tau \Delta v}\,, \\
\ell_v(v) & = \frac{D}{\tau^2 \Delta v^2} + \frac{v}{2 \tau \Delta v} \,.
\end{align}
\end{subequations}
In principle, these rates also suffer from possibly being negative, however, $\Delta v$ can always be minimally adjusted so that $r_v(v)$ and $\ell_v(v)$ vanish strictly before becoming negative, so that this region of phase space is never accessed.

Following the canonical procedure \cite{TaeuberHowardVollmayr-Lee:2005,Taeuber:2014,pausch_topics_2019} to write a master equation in terms of ladder operators, we introduce 
introduce creation and annihilation operators $a^{\dagger}_{v,x}$ and $a_{v,x}$  
\begin{align}
a^{\dagger}_{x,v} \bra{\{n\}} = \bra{n_{x,v}+1}, \qquad a_{x,v} \bra{\{n\}} = n_{x,v} \bra{n_{x,v}-1} \,  ,
\end{align}
where the pure states show only the occupation number subject to the operator.
With these states we can define the bra vector
\begin{align}
\bra{\psi}(t) = \sum_{\{n\}} p_t(\{n\}) \bra{\{n\}} \;,
\end{align}
whose time evolution 
\begin{align}
\der{}{t}\bra{\psi}(t) = \mathcal{H}[a,a^{\dagger}]  \bra{\psi}(t) \; ,
\end{align}
can now be written in terms of a "Hamiltonian operator" $\mathcal{H}[a,a^{\dagger}]$, which needs to be determined from the master equation, and thus from the rates \Erefs{rates_hopping} and \eref{rates_v}.
Replacing the factors of $n_{x,v}$ in the Master equation by suitable combinations of $a^{\dagger}_{x,v}$ and $a_{x,v}$ we find 
\begin{multline}
\mathcal{H}[a,a^{\dagger}] =  \sum_{v,x}  \Bigg\{ 
\frac{D_x}{\Delta x^2} a^{\dagger}_{x,v}(a_{x+\Delta x,v}-2 a_{x,v}+a_{x-\Delta x,v}) 
+ \frac{\theta(v) v a^{\dagger}_{x,v}}{\Delta x} (a_{x-\Delta x,v} -a_{x,v}) 
- \frac{\theta(-v) v a^{\dagger}_{x,v}}{\Delta x} (a_{x+\Delta x,v} -a_{x,v})  \\
+\frac{D}{\tau^2 \Delta v^2} a^{\dagger}_{x,v}(a_{x,v+\Delta v}-2 a_{x,v}+a_{x,v-\Delta v}) 
+ \frac{a^{\dagger}_{x,v}}{2\tau\Delta v} ((v+\Delta v) a_{x,v+\Delta v} -(v-\Delta v)a_{x,v-\Delta v})
\Bigg\} \,.
\end{multline}
Since this Hamiltonian is already normal ordered we can substitute the operators by the fields $a^{\dagger}_{x,v} \to \field^\dagger(x,v,t)$ and $a_{x,v} \to \field(x,v,t)$ and take the continuum limit, as we would have done in the master \Eref{master}. 
Finally, a Doi-Shift $\field^{\dagger} = \rfield+1$ \cite{Cardy:2006} simplifies the diagrammatics drastically and comes here at no price, as say $\sum_x a_{x,v}-a_{x+\Delta x,v}$ simply telescopes,
we finally have
\begin{align}
\elabel{hamiltonian}
\mathcal{H}[\field, \rfield] = 
\rfield \Bigg\{
D_x \npder{}{x}{2}
- v \pder{}{x} 
+ \frac{D}{\tau^2} \npder{}{v}{2}
+ \frac{1}{\tau} \pder{}{v} v \Bigg\} \field \,, 
\end{align}
which agrees exactly with the action \Eref{action}.




\end{document}

%% file: MB_commands.tex
\newcommand{\dbar}[1]{\ensuremath{\textrm{\dj}#1} \,}
\newcommand{\diff}[1]{\ensuremath{\textrm{d}#1}\,}
\newcommand{\Diff}[1]{\ensuremath{\mathcal{D} [ #1 ]}\,}

\newcommand{\der}[2]{\ensuremath{\dfrac{\text{d}#1}{\text{d}#2}}}

\newcommand{\pder}[2]{\ensuremath{\dfrac{\partial#1}{\partial#2}}}
\newcommand{\npder}[3]{\ensuremath{\dfrac{\partial^{#3}#1}{\partial#2^{#3}}}}

\newcommand{\ve}[1]{\vec{#1}}
\newcommand{\bra}[1]{| #1 \rangle }

\newcommand{\EXP}[1]{\operatorname{exp}\!\Bigg\{#1\Bigg\}}

\renewcommand{\exp}[1]{\mathchoice{\mathrm{e}^{#1}}{\operatorname{exp}\left(#1\right)}{\operatorname{exp}\left(#1\right)}{\operatorname{exp}\left(#1\right)}}

\newcommand{\mean}[1]{\ensuremath{\left\langle\,#1\,\right\rangle}}

\def\deltabar{{\mathchar '26\mkern -10mu\delta}}
